\documentclass[
preprint,
 amsmath,amssymb,
 aps,
]{revtex4-2}

\usepackage{graphicx}
\usepackage{dcolumn}
\usepackage{bm}

\usepackage{siunitx}

\begin{document}

\title{Sedimentation of a single soluble particle\\at low Reynolds and high Péclet numbers}

\author{Nan He}
\email{nan.he@espci.fr}
\author{Yutong Cui}%
\author{David Wai Quan Chin}%
\author{Thierry Darnige}%
\author{Philippe Claudin}%
\author{Beno\^{i}t Semin}%
 \email{benoit.semin@espci.fr}
\affiliation{%
 PMMH, CNRS, ESPCI Paris, Universit\'e PSL, Sorbonne Université, Universit\'e Paris Cit\'e, F-75005, Paris, France
}%

\date{\today}

\begin{abstract}
We investigate experimentally the dissolution of an almost spherical butyramide particle during its sedimentation, in the low Reynolds high P\'eclet regime. The particle sediments in a quiescent aqueous solution, and its shape and position are measured simultaneously by a camera attached to a translation stage. The particle is tracked in real time, and the translation stage moves accordingly to keep the particle in the field of the camera. The measurements from the particle image show that the radius shrinking rate is constant with time, and independent of the initial radius of the particle. We explain this with a simple model, based on the sedimentation law in the Stokes' regime and the mass transfer rate at low Reynolds and high P\'eclet numbers. The theoretical and experimental results are consistent within $20\%$. We introduce two correction factors to take into account the non-sphericity of the particle and the inclusions of air bubbles inside the particle, and reach quantitative agreement. With these corrections, the indirect measurement of the radius shrinking rate deduced from the position measurement is also in agreement with the model. We discuss other correction factors, and explain why there are negligible in the present experiment. We also compute the effective Sherwood number as a function of an effective P\'eclet number.
\end{abstract}

\maketitle

\section{Introduction}
The mass transfer from a solid soluble particle in a fluid is of major relevance in chemical engineering \citep{McCabe1993}, for example in food industry \citep{Fryer1997} and in pharmaceutical industry \citep{grijseelsHydrodynamicApproachDissolution1981,Dokoumetzidis2006}. Mass transfer from particles associated to phase change also occurs in a geophysical context, for instance the melting of rocks in magma \citep{mcleod_riley_sparks_1996}, the sublimation of ice drops in the atmosphere \citep{chouippeHeatWaterVapor2019}, or the melting of snow and hail when reaching a sea \citep{vahab_murphy_shoele_2022}.

The dynamics of dissolution is different for isolated particle and for a suspension of particles. In the latter case, the particles interact through the concentration of the solute in the fluid phase, and due to the hydrodynamic interaction between the particles \citep{Kriaa2022effectsa}. Moreover, the dissolution of the particle can affect the density of the fluid phase, inducing buoyancy effects like the formation of plumes. In the present article we focus on the case of an isolated particle.

All these rather different configurations correspond to different regimes, which have been classified following the two main dimensionless parameters which control the mass transfer from a single spherical particle. A usual choice (which will be ours, see next section) is to use the Reynolds $\mathrm{Re}$ and the P\'eclet $\mathrm{Pe}$ numbers. An alternative to $\mathrm{Pe}$ is to use the Schmidt number $\mathrm{Sc}$, where these three numbers are related by $\mathrm{Pe}= \mathrm{Re} \times \mathrm{Sc}$. Both $\mathrm{Re}$ and $\mathrm{Pe}$ vanish for a motionless particle, for which the mass transfer is only due to molecular diffusion. In gases, these two numbers are usually similar, whereas in liquids the P\'eclet number is several orders of magnitudes higher than the Reynolds number. The dimensionless mass transfer is the Sherwood number $\mathrm{Sh}$, defined here (see Eq.~\ref{sherwood} below) as the ratio of the total mass transfer by its purely diffusive component. Many past studies have performed experiments or numerical simulations, or developed models, to relate $\mathrm{Sh}$ to $\mathrm{Re}$ and $\mathrm{Pe}$. Finally, in order to better cover the subject, it is also important to consider heat transfer from a particle, which is analogous to mass transfer within some hypotheses, in particular negligible radiation. In the case of heat transfer, the analog to $\mathrm{Pe}$ is the thermal P\'eclet number defined with the thermal diffusivity, the equivalent to $\mathrm{Sc}$ is the Prandtl number $\mathrm{Pr}$, and the equivalent to $\mathrm{Sh}$ is the Nusselt number $\mathrm{Nu}$.

The first regime is when both $\mathrm{Re}$ and $\mathrm{Pe}$ are small. It has for example been investigated experimentally by \cite{So2022}, measuring the size of an almost spherical succinic acid particle in unstirred water as a function of time. The results are compatible with a purely diffusive mass transfer (i.e. $\mathrm{Sh} = 1$ with our definition). The diffusive mass transfer has also been verified experimentally in the case of a droplet of hexadecane levitated in an eletrodynamic balance and undergoing a flow of N$_2$ and helium \citep{zhangMassTransferSingle1987}. Since the fluids are at rest, the expected result is similar for a liquid or a solid particle in this regime.

The regime we are particularly interested in this work is the case of small $\mathrm{Re}$ and large $\mathrm{Pe}$, which, as we already said, is possible in a liquid. A fundamental analytical calculation for a sphere has been performed by \cite{levichPhysicochemicalHydrodynamics1963}, which gives $\mathrm{Sh} \propto \mathrm{Pe}^{1/3}$. This scaling law has been confirmed by numerical simulations \citep{Clift1978, Assuncao2023} as well as experiments using two rotating cylinders to impose an homogeneous flow and electrochemical measurements to obtain the mass flux \citep{kutateladze_nakoryakov_iskakov_1982}. A similar configuration is the sinking of small spheres in a turbulent flow, which has been investigated experimentally and numerically by \cite{lawson_ganapathisubramani_2023}.

Finally, many experiments and simulation have been performed in the regime where both $\mathrm{Re}$ and $\mathrm{Pe}$ are large, for a fixed particle submitted to a uniform flow or a free falling particle. In this case, one expects the scaling law $\mathrm{Sh} \propto \mathrm{Re}^{1/2} \mathrm{Sc}^{1/3}$ \citep{Lochiel1964}, and most of the results are effectively compatible with a correlation $\mathrm{Sh} \propto \mathrm{Re}^{1/2} \mathrm{Sc}^\alpha$, $\alpha$ is in the range $0.3$--$0.4$, plus correction terms. Using numerous simulations, \cite{Melissari2005} have found $\alpha=0.36$ in the case of heat transfer for $3\times10^{-3} < \mathrm{Pr} < 10^1$ and   $10^2 < \mathrm{Re} <  5 \times 10^4$. The experiments include various systems: sedimentation and dissolution of urea spheres in a vertical glass column \citep{Petrescu1997}, dissolution of benzoic acid spheres in a flow (natural or forced convection) of water or propylene glycol \citep{steinbergerMassTransferSolid1960}, dissolution of ice ball in a hydrodynamic channel with water flow \citep{haoHeatTransferCharacteristics2002}, dissolution of hard candy submitted to a flow \citep{huangShapeDynamicsScaling2015}. However, some other experimental configurations with a very different flow, like large ice balls melting in a turbulent von K\'arm\'an flow \citep{machicoaneMeltingDynamicsLarge2013}, leads to different scalings.

Spherical versus non-spherical particles have also been studied in this large $\mathrm{Re}$-$\mathrm{Pe}$ regime. One can in particular mention experiments on the dissolution of neutrally buoyant particles with rectangular cuboid initial shapes in isotropic turbulent environments \citep{oehmkeNewParticleMeasuring2021}. Numerically, simulations of heat transfer past spheres, cuboids and ellipsoids been performed by \cite{richterDragForcesHeat2012}, and empirical correlation for non-spherical ellipsoids have been obtained by \cite{Kishore2011,Ke2018,Chen2021,Kiwitt2022}. Finally, several correction factor to the mass transfer, associated with the aspect ratio of non-spherical particles, the thermal effect due to dissolution, and the finite solubility effect have been discussed in \citep{Elperin2001,Assuncao2023}.

In the present work, we investigate the dissolution of an almost spherical particle that sediments in an aqueous solution at rest. These experiments belong to the low Reynolds and high P\'eclet regime, which, in comparison to the high Reynolds regime, has been less studied. In the next section 2, we present the theoretical framework that we need to interpret the results, showing in particular that we expect a constant reduction rate of the particle size. We describe the experimental setup in section 3 which allows us to measure simultaneously the position and the shape of the particle during its sedimentation. We emphasise the use of particles made of butyramide, a chemical which does not change the density of water when dissolving, hence preventing any buoyancy effect in the fluid. Section 4 is devoted to the comparison of our experimental results with the model, and we show that we can make it quantitative accounting for shape and density correction factors. Finally, conclusions and perspectives are drawn in section 5. Some technical aspects are gathered in supplementary material.

\section{Theory for a spherical particle}

We present in this section the theory of a spherical particle falling in a quiescent fluid. We focus on the regime for which  the particle Reynolds number $\mathrm{Re} = \rho_f a U/\eta$ is low, where $a$ is the particle radius, $U$ its settling velocity, $\eta$ is the dynamic viscosity of the fluid and $\rho_f$ its mass density. It means that the fluid motion around the particle can be described by the Stokes equations, where the fluid inertia can be neglected. The particle velocity then results from the balance of the three relevant forces: the downward force of gravity $F_g$, the upward buoyant force $F_b$, and the drag force $F_d$, which classically express as:
\begin{eqnarray}
F_g & = & \frac{4}{3}\pi a^3 \rho_p g,
\label{F_g} \\
F_b & = & \frac{4}{3}\pi a^3 \rho_f g,
\label{F_b} \\
F_d & = & 6\pi \eta a U.
\label{F_d}
\end{eqnarray}
$\rho_p$ is density of the particle and $g$ is gravity acceleration. The resulting settling velocity of the particle is:
\begin{equation}
U = \frac{2}{9} \, \frac{(\rho_p - \rho_f) g}{\eta} a^2 .
\label{U}
\end{equation}

This particle can dissolve in the fluid, and we assume that it does so in the regime where its P\'eclet number, defined as $\mathrm{Pe} = \frac{Ua}{D}$, where $D$ is the diffusion coefficient of diffusion of the dissolved matter composing the particle into the fluid, is large. Both small $\mathrm{Re}$ and large $\mathrm{Pe}$ are encountered for small soluble particles in liquids as one typically has $D \ll \eta/\rho_f$ (Parameters can be found in Table~\ref{table.1}). 

In this regime, following the analytical calculation of \cite{levichPhysicochemicalHydrodynamics1963}, the mass transfer rate $\dot{m}$ of the sphere, i.e. the mass the particle losses per unit time when is dissolves, can be expressed in terms of the Sherwood number as:
\begin{equation}
\mathrm{Sh} = - \frac{\dot{m}}{4\pi Dac_0} \simeq \frac{2}{\pi} \mathrm{Pe}^{1/3},
\label{sherwood}
\end{equation}
where the dot denotes time derivative, and $c_0$ is the concentration (in kg/m$^3$) of the dissolved matter that closes to the particle. In first approximation, $c_0$ is the saturated concentration of the solute. Relating the mass of the sphere to its radius $m = \frac{4}{3} \pi \rho_p a^3$, the mass transfer rate can also be written as:
\begin{equation}
\dot{m} = 4\pi \rho_p a^2 \dot{a}.
\label{dotM2}
\end{equation}
Equating (\ref{sherwood}) and (\ref{dotM2}), we obtain
\begin{equation}
\dot{a}= -\frac{2}{\pi} \left(\frac{2}{9} \right)^\frac{1}{3} \frac{D^\frac{2}{3} c_0}{\rho_p} \left(\frac{( \rho_p -  \rho_f ) g}{\eta}\right)^\frac{1}{3} .
\label{dota2}
\end{equation}
All factors on the right hand side of the above expression only depend on the characteristics of the fluid and the particle. The rate at which the particle size decreases over time is thus constant, i.e. independent of the radius of the particle, resulting in a linear relationship between $a$ and $t$. We will test this remarkably simple behaviour experimentally in the present article.

\section{Experimental set-up}

\begin{figure}[htbp]
    \centering
    \includegraphics[width = 0.65\textwidth]{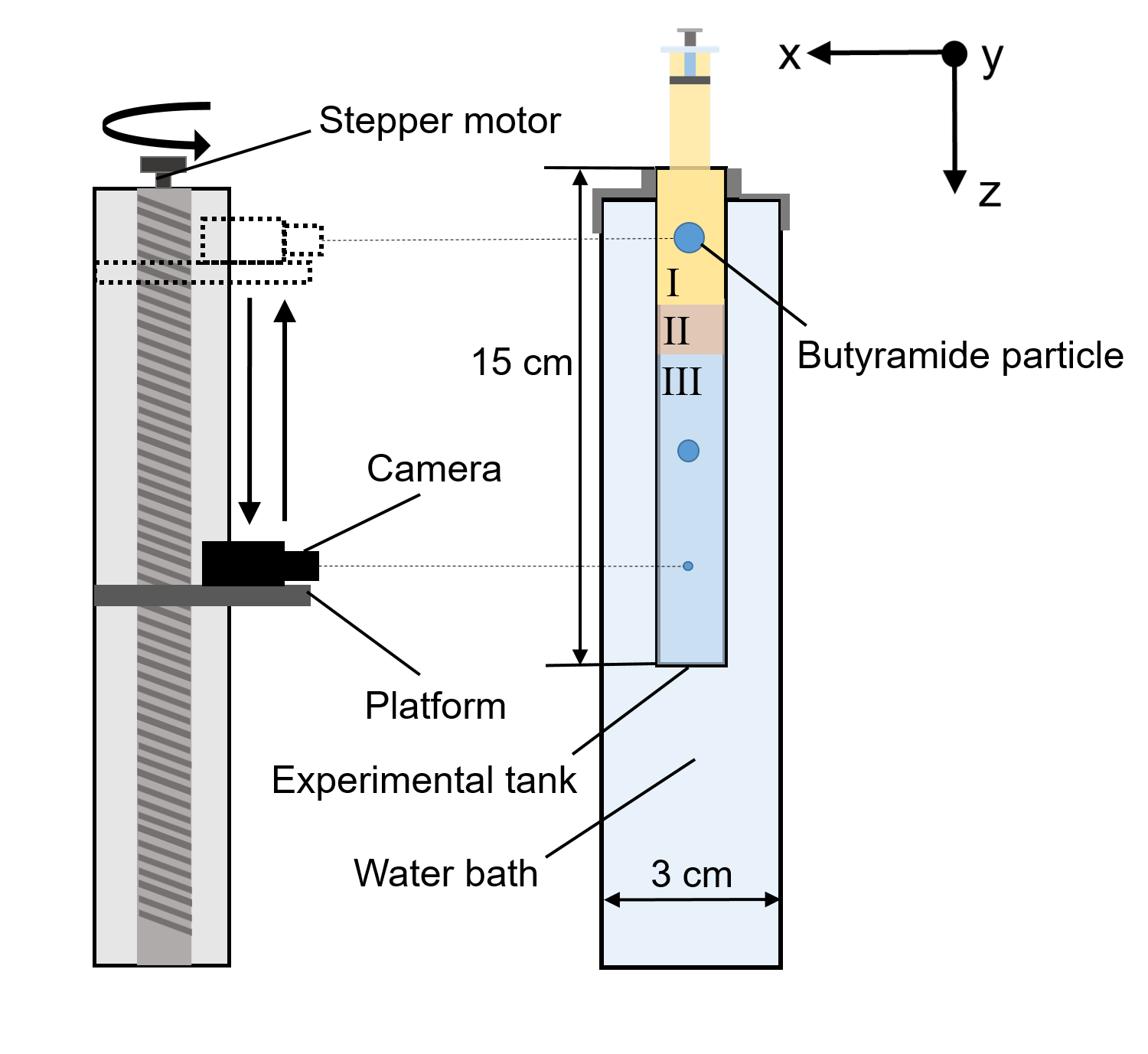}
    \caption{Schematic diagram of the experimental set-up. In the tank, the upper layer I is saturated butyramide and the lower layer III is the NaCl solution. The transition layer is numbered II.}
    \label{fig.exp}
\end{figure}

We have built an experiment to investigate the dynamics of such a particle that sediments and continuously dissolves, resulting in a reduction of its size and mass. The setup consists in an elongated tank and a particle tracking system, as illustrated in Figure \ref{fig.exp}. The tank has an inner width of \SI{10}{\milli\meter} and an inner length of \SI{150}{\milli\meter}. As the particles we consider are rather small (on the order of 100~$\mu$m), and thus easily disturbed by small velocity fluctuations, the tank is placed in a larger water bath (internal width of \SI{30}{\milli\meter}) to avoid convective disturbances inevitably caused by small temperature differences between the two sides of the experiment. The transparency of the two tanks allows for visual observation of the particle sedimentation process. The reliability of the entire experimental device has been verified through the sedimentation of plastic beads in distilled water. The two tanks are joined at the top by a detachable connector to ensure the verticality of the inner tank and prevent temperature fluctuations in the water bath caused by evaporation. In the present study, a camera with a resolution of $1936\times1216$ pixels, manufactured by IDS industrial camera company, was used. Prior to the experiments, careful scale calibration was performed, with a typical resolution of 5 pixels/\SI{}{\micro\meter}. The particle tracking was controlled by a self-written Labview program, inspired by \cite{Darnige2017lagrangian}.  The camera was connected to a computer to measure in real-time the position of the particle in the image, enabling to move the linear stage from PI (Physical Instrument) and thus the camera to follow the particle. Thus, synchronous position information and images of the particle were obtained from the tracking system.

The density of the aqueous solution containing the dissolved matter from the solid particle is usually larger than that of the pure water -- this is the case for NaCl for example. Here we use particles of butyramide, a chemical whose saturated solution has a density very close to that of pure water, which minimise the effect of density increase around the particle during its sedimentation-dissolution motion. At \SI{21}{\celsius}, the typical temperature at which experiments were run, we indeed measured with a pycnometer and a precision scale that the density of the saturated butyramide is $0.998 \pm 0.001$  \SI{}{\gram\per\cubic\centi\meter}, i.e. similar to the density of water at the same temperature (see Table~\ref{table.1}). Butyramide is also very soluble in water \citep{romeroSolubilityAcetamidePropionamide2010a}, and the crystals have a bulk density of \SI{1.032}{\gram\per\cubic\centi\meter}.
We made the particles using a stainless steel tip with an inner diameter of around 1 mm, resulting in particles with a shape similar to a cylinder with an aspect ratio close to unity.

Before the experiment is started, two different layers are prepared in the tank. The upper layer is a saturated butyramide solution in which the particle cannot dissolve. The lower layer is water with some dissolved NaCl, in order to make it slightly heavier than the upper layer for stability. The amount of NaCl is such that the density of this lower layer is \SI{1.014} {\gram\per\cubic\centi\meter}. We assume that this moderate presence of NaCl does not influence the dissolution process of butyramide in water. Once prepared, the two solutions are first put in a vacuum pump to remove air bubbles before being placed in the experimental tank. At the beginning of the experiment, pure water at room temperature is injected into the water bath. The sedimentation tank is then carefully placed and attached by the detachable connector. After the NaCl solution is poured at the bottom of the tank, the saturated butyramide is carefully injected using a syringe with a small-sized tip, providing a more stable flow. This process results in a narrow transition layer, and the interface between the upper and lower layers can be visually distinguished. The vertical thickness of the lower layer solution is \SI{95}{\milli\meter}, and that of the upper layer solution is \SI{45}{\milli\meter}. Butyramide particles that have been pre-stocked in a saturated solution are drawn into a syringe without a tip, and the syringe is then placed vertically on top of the tank. At this point, the saturated solutions in the syringe and in the tank are connected and the particles can start to sediment. Once a particle is in the field of view of the camera, the tracking system captures it immediately, and track it until its size becomes smaller than $\simeq$\SI{3}{\micro\meter}.

Moreover, the different physical parameters involved in this experiment have been carefully measured or determined, the value of them shown in Table \ref{table.1}. The dynamic viscosities of the saturated butyramide and NaCl solutions, respectively denoted $\eta_b$ and $\eta_n$, were measured with high precision using a rheometer Anton Paar specifically for liquids with a viscosity similar to water. The saturated butyramide concentration $c_0$ is well calibrated as a function of temperature in \cite{romeroSolubilityAcetamidePropionamide2010a}. For the diffusivity $D$, we recorded with a camera the refraction angle of the interface of a stratified solution consisting of a saturated butyramide solution and still water over time. $D$ was deduced from the square relationship between maximum reflected angle and time. More details about these measurement can be found in the supplementary material.

\begin{table}[htbp]
    \centering
    \begin{tabular}{|c|c|c|c|c|c|c|c|}
    \hline
        Particle & \multicolumn{4}{|c|}{Saturated butyramide solution}& \multicolumn{2}{|c|}{NaCl solution} & Water\\\hline
      $\rho_p$  & $D$ & $c_0$ &   $\rho_f$ & $\eta$ &$\rho_f$ &  $\eta_f$ & $\rho_f$ \\\hline
       $\SI{}{\gram\per\cubic\centi\meter}$ &\SI{}{\square\meter\per\second}& $\SI{}{\gram\per\cubic\centi\meter}$ &  $\SI{}{\gram\per\cubic\centi\meter}$ & mPa s & $\SI{}{\gram\per\cubic\centi\meter}$ &  mPa s &$\SI{}{\gram\per\cubic\centi\meter}$ \\\hline
      1.032 & $7.2 \times 10^{-10}$ & 0.182 & 0.998 & 1.853 & 1.014 &  1.018 & 0.9980\\\hline
       $  $ & $\pm 0.4 \times 10^{-10}$ & $\pm 0.005$ & $\pm 0.001$ & $\pm 0.002$ & $\pm 0.001$ &  $\pm 0.006$ & $\pm 0.0001$ \\\hline
    \end{tabular}
    \caption{Parameters of the experiment at $21 \pm 0.5$ \SI{}{\celsius}. The column about water is given for comparison. }
    \label{table.1}
\end{table}

\section{Results and discussion}
\label{sec:ResultsDiscussion}

\subsection{Simultaneous measurements of radius and position of the particle}
\begin{figure}[htbp]
    \centering
    \includegraphics[width = 1\textwidth]{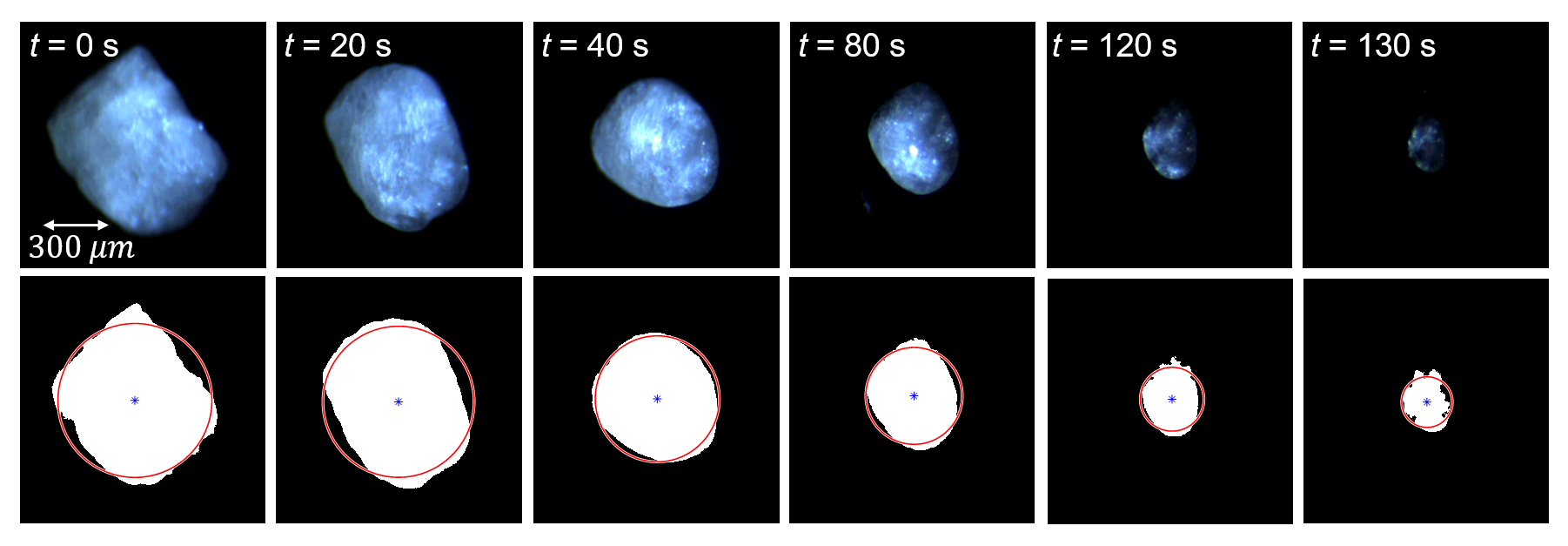}
    \caption{Dissolving process of the particle over time. The top line shows the original pictures, as captured by the camera during the experiment. The bottom line shows the corresponding processed images, where the white area $A$, obtained after convex closure of the binarised picture, represents the projected surface of the particle. On each of these bottom images, the blue-star point indicates the centroid of the white area, and the red circle, centred on that point, has the same surface as the white area, i.e. gives the equivalent radius $a = \sqrt{A/\pi}$ of the particle.}
    \label{fig.particleD}
\end{figure}

\begin{figure}[htbp]
     \centering
     \includegraphics[width = 0.6\textwidth]{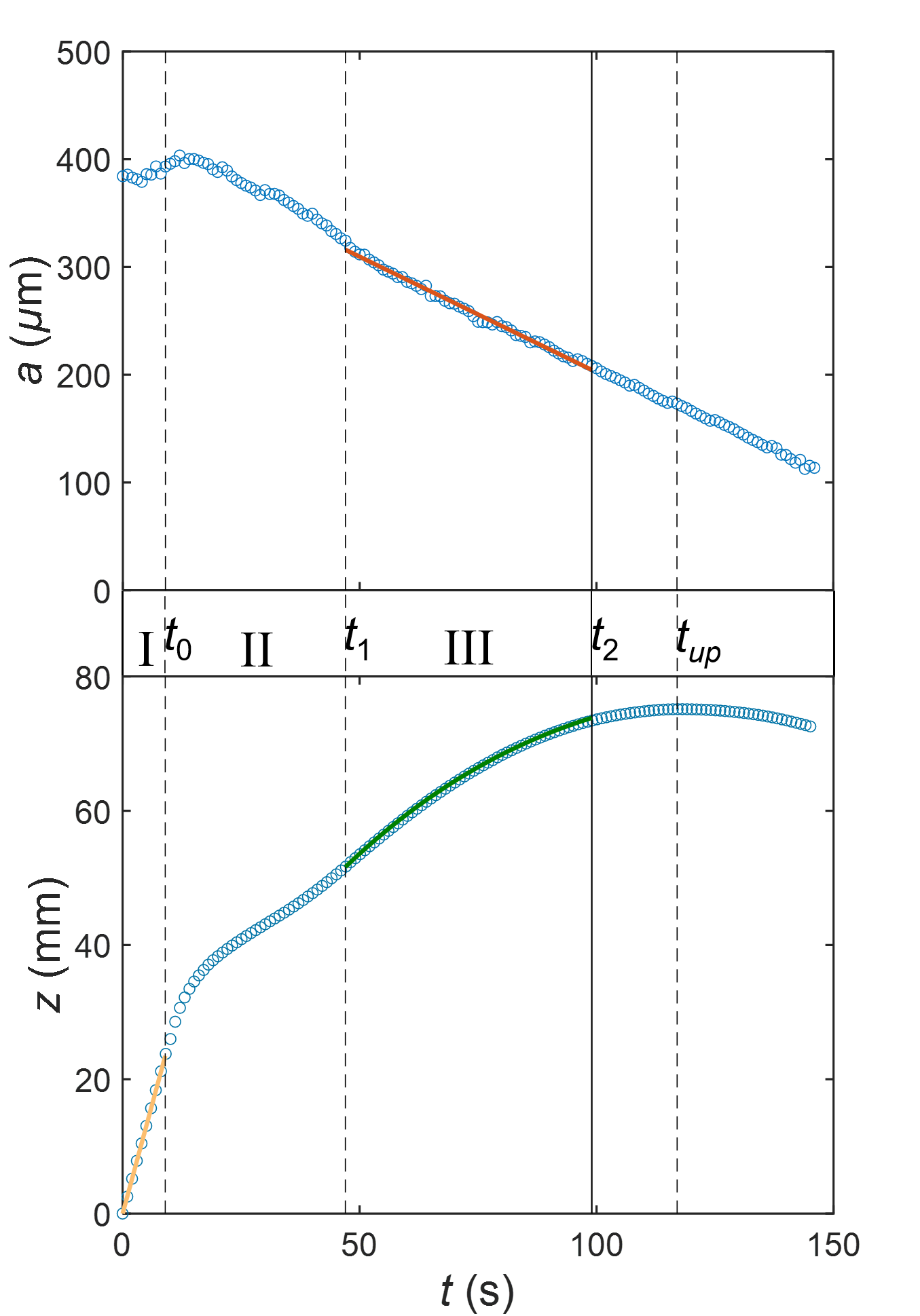}
     \caption{Time evolution of the equivalent radius $a$ (a) and the vertical displacement $z$ (b) of the particle. Blue circles: experimental data. The vertical dashed lines show the times separating the different stages: $t_0$ end of the saturated layer, $t_1$ end of the transition layer, $t_{\rm up}$ motion reversal. Those times were determined using the velocity deduced from the derivative of the displacement (b), see supplementary material. Solid black line: time before which the volume of the air bubbles attached to the particle is less than $1$\%. Red line: linear fit in stage III of the radius decrease to deduce $\dot{a}_1$. Yellow line: linear fit in stage I of the particle position to deduce $U_0$. Green curve: fit of $z(t)$ in stage III with Eq.~\ref{Displacement2}, from which another estimate $\dot{a}_2$ of the radius shrinking rate is obtained.}
     \label{fig:radius and displacement}
\end{figure}

Figure \ref{fig.particleD} illustrates the dissolution process of the particle during its sedimentation. We observe that the particle shrinks over time, gradually rounding off into a shape slightly elongated in the vertical direction. We can also notice a slight rotation of the particle. More quantitatively, we estimated the volume change of the particle by image analysis: binarising the picture of the particle with a suitable gray threshold and finding boundaries after convex hull, which fills the holes inside of binarised image, we could extract a projected area of the particle. From that surface, a centroid, which we take as the effective location $z$ of the particle, and an equivalent radius $a$ can be defined (Figure \ref{fig.particleD}, bottom line). Following these quantities picture after picture, we could this way measure $z$ and $a$ as functions of time, as displayed in Figure \ref{fig:radius and displacement}. 

After the particle is released from the syringe, it first sediments in the upper layer composed of a saturated butyramide solution. During this period (stage I), as the particle does not dissolve, its equivalent radius remains constant. Slight fluctuations can however be observed, caused by the rotation of the particle. By taking the average particle radius during this stage, the initial particle radius, denoted as $a_0$, can be obtained. The settling velocity of the particle also remains constant, and an initial value $U_0$ can be similarly computed from the average slope of the particle vertical displacement $z(t)$.

At time $t=t_0$ the particle enters the transition layer where the upper-layer butyramide and lower-layer NaCl solutions are mixing. This is a rather thin transition layer, but its stratification causes a significant drop of the particle settling velocity associated with an enhanced drag \citep{Magnaudet2020}. During this stage II, the particle begins to dissolve, and it does so with an almost constant radius shrinking rate. At time $t=t_1$ the particle has reached the lower layer and its velocity is back to normal settling values. As shown in Figure \ref{fig:radius and displacement}(a), throughout its sedimentation in this layer (stage III) the particle equivalent radius continues to decrease at a constant rate, which is consistent with the theoretical expectation (\ref{dota2}). A linear fitting of the data $a(t)$ gives a direct measurement of the radius shrinking rate, denoted as $\dot{a}_1$. We shall see below that this rate can be also estimated in another way.

Simultaneously, the particle velocity continuously decreases and notably reaches zero at some time $t_{\rm up}$, after which the particle motion is reversed, see Figure \ref{fig:radius and displacement}(b). This is due to some air bubbles trapped inside the particle during its preparation. As we detail in the following analysis, we will need to account for the fact that the effective density of the particle must be corrected by a factor $\beta_b$, associated with the presence of these bubbles. As the density difference $\rho_p - \rho_f$ is small, even a value of $\beta_b$ close to unity has a significant quantitative effect. Of course, such a constant correction factor cannot reproduce the particle motion reversal. Instead, close to that moment, the volume of the bubbles $V_b$ inside the particle can be assumed constant, so that, as the particle matter further dissolves, its effective density becomes less than that of the surrounding fluid solution and it eventually rises. One can one then can compute $V_b$ at that reversing time with 
\begin{equation}
V_b \left( \rho_p - \rho_g \right) =  \frac{4}{3} \pi  \left(\rho_p - \rho_f \right)a^3(t = t_{\rm up}),
\label{VolumeBubbletup}
\end{equation}
where $\rho_g = 1.2$~kg/m$^3$ is the air density. Notice that, interestingly, the value of the rate $\dot{a}$ remains unchanged during the rising stage. We define the time $t_2$ at which $V_b$ represents $1$\% of the overall particle volume. Later analysis will then be restricted to times between $t_1$ and $t_2$, so that the effect of these bubbles in the particle sedimentation is small.

The radius shrinking rate can be alternatively obtained from the particle position. As $\dot{a}$ is a constant (Eq.~\ref{dota2}), $a$ in the expression of the settling velocity $U$ (Eq.~\ref{U}) can then be replaced by $a=a_1+\dot{a}(t-t_1)$, where $a_1 = a(t_1)$. Integrated once, the vertical position of the particle thus writes
\begin{equation}
z = \int_{t_1}^t U dt' =  \frac{2}{9} \, \frac{( \rho_p -\rho_f ) g }{\eta} \left[ \frac{1}{3}\dot{a}^2 (t-t_1)^3 + a_1 \dot{a} (t-t_1)^2 + a_1^2 (t-t_1) \right] + z_1,
\label{Displacement2}
\end{equation}
where the $z_1$ is particle vertical position at the beginning of stage III (time $t_1$). Fitting this expression to the data $z(t)$ allows us to get a value of $\dot{a}$, which we denote as $\dot{a}_2$ to distinguish with the more direct estimate $\dot{a}_1$. Importantly, in this fitting process, two other parameters are determined by fit: $a_1$ and $z_1$. The precision on fitting parameters is good (the maximal error bar for $\dot{a}_2$ is $\pm 0.05$ \SI{}{\micro\meter\per\second}), and we have for instance checked that imposing $a_1$ from the measurement $a(t)$ in the fitting of $z(t)$ leads to consistent results.

\subsection{Radius shrinking rate $\dot{a}$}
\begin{figure}[htbp]
    \centering
    \includegraphics[width = 0.6\textwidth]{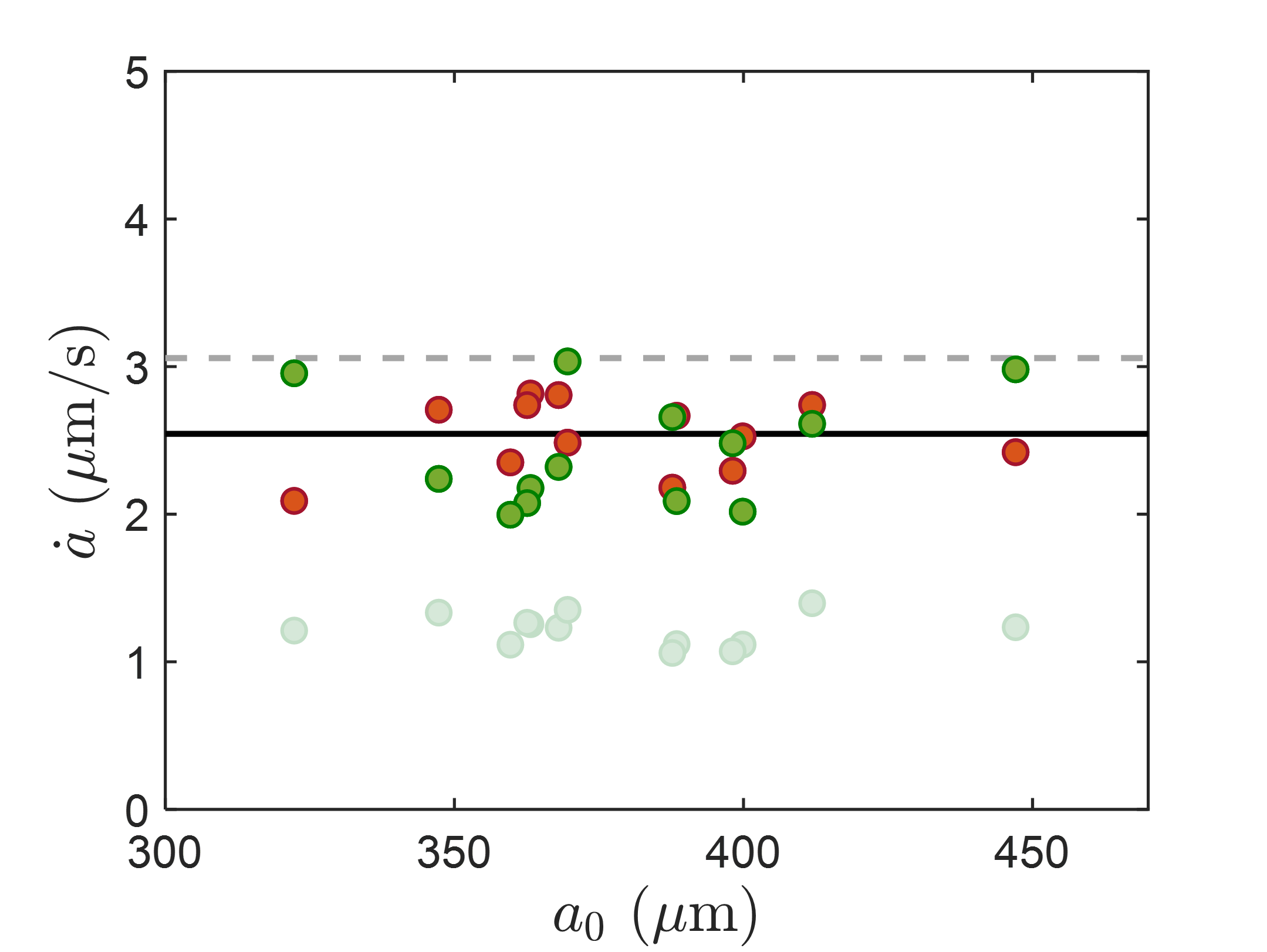}
    \caption{Reduction rate of the particle radius $\dot{a}$ from theory and data analyses for various initial particle size $a_0$. Red dots: direct measurement from the linear fitting of $a(t)$ (see Fig.~\ref{fig:radius and displacement}a). These are what we denoted as $\dot{a}_1$. Green dots: indirect values obtained from the fitting of $z(t)$ (see Fig.~ \ref{fig:radius and displacement}b). These are what we denoted as $\dot{a}_2$. Light green points: fitting without any correction factors, i.e. using Eq.~\ref{Displacement2}. Dark green points: fitting accounting for correction factors, i.e. using Eq.~\ref{Displacement2corrected}. Gray dash line: uncorrected theory (\ref{dota2}). Black solid line: corrected theory (\ref{dotacorrected}). Data dispersion shows the overall precision we can reach, but from the fitting process of a single experimental run, errors bars are smaller than the symbol size.}
    \label{fig:result}
\end{figure}

As theoretically expected and showed in Figure~\ref{fig:result}, we find both $\dot{a}_1$ and $\dot{a}_2$ constant, i.e. independent of the initial size $a_0$ of the particle. The prediction from (\ref{dota2}) is above the $\dot{a}_1$ measurements by 20\%. Moreover, although on the same order, $\dot{a}_2$ is systematically smaller than $\dot{a}_1$ by a factor of $\simeq 2$. These discrepancies prompt us to revisit the above theoretical expressions in order to understand where the idealised case of a homogeneous spherical particle we have considered must be corrected. We have already mentioned in the previous section that the presence of trapped air bubbles must be accounted for with an effective particle density corrected by a factor $\beta_b$. Another important aspect is the geometry of the particle. Since we have only access to a projection of the particle shape, it is unlikely that the effective radius $a$ we have introduced quantitatively works for the particle volume. This volume is key for the computation of the gravity and buoyancy forces. To account for this volume uncertainty, we introduce a correction factor $\beta_a$ that will multiply the radius in the expression of these forces. We will discuss later in section~\ref{Correction factors} why we do not introduce correction factors for the other variables. With these two correction factors $\beta_a$ and $\beta_b$, the expressions for $U$, $\dot{a}$ and $z$ rewrite
\begin{eqnarray}
U & = & \frac{2}{9} \, \frac{(\beta_b \rho_p - \rho_f) g}{\eta} \beta_a^3 a^2,
\label{Ucorrected} \\
\dot{a} & = & -\frac{2}{\pi} \left(\frac{2}{9} \right)^\frac{1}{3} \frac{D^\frac{2}{3} c_0}{\beta_a^2 \beta_b \rho_p} \left(\frac{( \beta_b \rho_p -  \rho_f ) g}{\eta}\right)^\frac{1}{3},
\label{dotacorrected} \\
z & = & \int_{t_1}^t U dt' =  \frac{2}{9} \, \frac{( \beta_b \rho_p -\rho_f ) g }{\eta} \beta_a^3 \left[ \frac{1}{3}\dot{a}^2 (t-t_1)^3 + a_1 \dot{a} (t-t_1)^2 + a_1^2 (t-t_1) \right] + z_1.
\label{Displacement2corrected}
\end{eqnarray}

Using the data in stage I (upper layer), where $U_0$ and $a_0$ are measured accurately and for which the density as well as the viscosity of saturated butyramide are known, (\ref{Ucorrected}) gives a first relationship between the correction factors $\beta_a$ and $\beta_b$. Similarly, with the linear fit of the radius reduction in stage III (lower layer) giving the rate $\dot{a}_1$, layer in which the density, the viscosity as well as the diffusivity of NaCl solution are known, (\ref{dotacorrected}) gives a second relationship linking $\beta_a$ and $\beta_b$. They can be solved numerically, and, upon ensemble averaging over 13 independent experimental runs, we obtained $\beta_a = 0.921 \pm 0.002$ and $\beta_b = 0.988 \pm 0.002$. With these values, the fit of the curve $z(t)$ in stage III with (\ref{Displacement2corrected}) allows us to deduce a new value of $\dot{a}_2$. As shown in Figure~\ref{fig:result}, the theoretical prediction of $\dot{a}$ now fits the direct measurements $\dot{a}_1$ as it should, and the corrected $\dot{a}_2$ are now quantitatively consistent with $\dot{a}_1$. Importantly, these corrections assume that these factors can be taken constant over the whole sedimentation process (in fact, until time $t_2$).

A value of $\beta_b$ so close to unity may seem surprising, but because we are dealing with a small density difference between particle and fluid, these numerical adjustments are very sensitive. In fact, trying to impose $\beta_b=1$, we were not able to reach a quantitative matching of $\dot{a}_1$, $\dot{a}_2$ and theory as in Figure~\ref{fig:result} playing with $\beta_a$ only. Furthermore, the value we got for $\beta_a$ corresponds to an actual volume of the particle about $3/4$ times smaller than deduced from the surface-induced effective radius $a$. This is consistent with particles in the form of an ellipsoid with its major axis parallel to the vertical, as observed in the experiments (see supplementary material).

\subsection{Other correction factors}
\label{Correction factors}

Other correcting factors could of course be considered. First of all, if ellipsoid-shaped particles are at play, the drag force is modified by the particle aspect ratio $E$, defined as the ratio between the major and the minor axes lengths. Following the work of \cite{lothDragNonsphericalSolid2008} for spheroids, the Stokes drag correction factor for a motion parallel to the major axis follows the relation
\begin{equation}
\beta_{\rm drag} = \left( \frac{4}{5} + \frac{E}{5} \right) E^{-1/3}.
\label{betadrag}
\end{equation}
The analysis of the pictures of the sedimenting particles shows aspect ratios typically between $1$ and $2$, with an average around $1.3$ (see supplementary material). This corresponds to a drag correction factor $\beta_{\rm drag} \simeq 0.97$. Accounting for it in the force balance, the above analysis of the particle's dynamics is not significantly affected: variations by less than a percent are found for $\beta_a$ and $\beta_b$, and around $1$\% for $\dot{a}$. At the first order, this shape effect on the drag can then be ignored for the present problem.

The shape of the particle influences its mass and heat transfer processes as well. The heat transfer from a particle, which is analogous to mass transfer within some hypotheses, in particular negligible radiation. We use the work of \cite{Chen2021} that provides Nusselt numbers for ellipsoids across a wide range of aspect ratios, $E$. Here with Re $\simeq 0.1$ and $\mathrm{Sc} \simeq 1400$, which are typical values of these experiments, we obtain $\beta_m \simeq 1.008$ for the $\mathrm{Sh}$ ratio between $E = 1.3$ and $E = 1$. Including this correction factor into the theoretical framework has a negligible impact on the results. As demonstrated in the previous paragraph, deviations of less than one percent are noted for $\beta_a$, $\beta_b$ and $\dot{a}$. 



The theory of \cite{levichPhysicochemicalHydrodynamics1963} which gives equation~(\ref{sherwood}) for the value of the Sherwood number rely on the hypotheses that the concentration of the solute is infinitesimal, and that there is no thermal effects during the dissolution. These two hypotheses are not verified for butyramide \citep{romeroSolubilityAcetamidePropionamide2010a}: the solubility of butyramide is large, and its dissolution in water is endothermic. In the following we evaluate the corresponding correction factor $\beta_{\rm sol}$ using the results of \cite{Elperin2001}.
These results are valid in the case of high P\'eclet and Schmidt number, which is the regime of the present experiments. The Sherwood number of equation~(\ref{sherwood}) is modified by a factor
\begin{equation}
\beta_{\rm sol} = \frac{1}{\gamma^{-1}-J^{2/3}/K}.
\label{betaS}
\end{equation}
In this expression $\gamma$ is a correction factor introduced by \cite{Elperin2001} which depends on the weight fraction of the solute at solid-liquid interface and in the bulk of the liquid. We estimate $\gamma = 1.09$ for butyramide using a linear fit from the data of table 2 of \cite{Elperin2001}.
$J$ is ratio of the molecular $D$ of the solute by the thermal diffusivity of the liquid $\alpha$. The value of $\alpha$ for water is $\alpha = 1.45 \times 10^{-7}$ 
\SI{}{\meter\squared\per\second} \citep{Balderas-Lopez2000thermalwave}. $K$ is a dimensionless number involving the specific heat of the liquid $c_p$, the latent heat of absorption $L$ and a coefficient $d$, which is the slope of the relationship of the concentration and temperature.  For water, the specific heat is $c_p =$  \SI{4.15}{\kilo\joule\per\kilo\gram\per\kelvin},  $d=0.01$~\SI{ }{\per\kelvin} and $L = $ \SI{-400}{\kilo\joule\per\kilo\gram} was sourced from the measurements by \citep{romeroSolubilityAcetamidePropionamide2010a}. The positive value of $d$ implies that heat absorbed during butyramide dissolution results in a decrease of the interfacial temperature and equilibrium concentration. 
The value of correction factor $\beta_{\rm sol} $ experiences only a minor change when the two values evaluated based on water are replaced by those calculated using NaCl solution: $\beta_{\rm sol} $ changes from 1.057 to 1.062, so that we take $\beta_{\rm sol} \simeq 1.06$.
Incorporating this value into the theoretical analysis does not affect much the results, with, as in the above paragraph, variations by less than a percent are found for $\beta_a$ and $\beta_b$, and around $1$\% for $\dot{a}$. This correction can thus be neglected at first order for the present analysis.

\subsection{Effective P\'eclet and Sherwood numbers}

\begin{figure}[htbp]
    \centering
    \includegraphics[width = 0.6\textwidth]{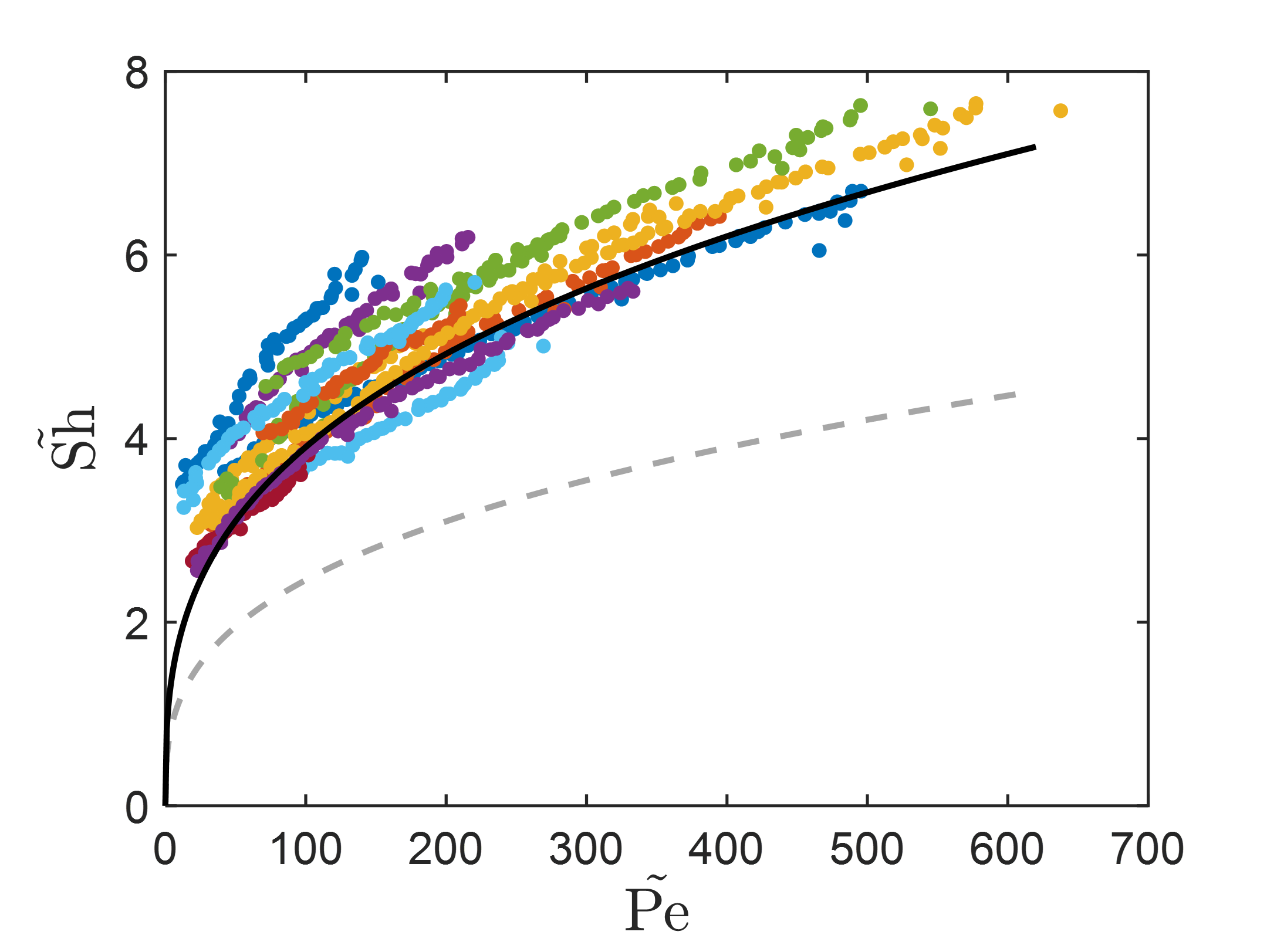}
    \caption{Effective Sherwood number vs effective P\'eclet number. Symbols: experimental data, corrected by their factors $\beta_a$ and $\beta_b$ computed as explained in the text after equations~(\ref{Ucorrected}-\ref{Displacement2corrected}). The colours correspond to different runs. Gray dash line: uncorrected theory. Black solid line: theory accounting for averaged correction factors.}
    \label{fig:shversuspe}
\end{figure}

These experimental data finally allow us to assess the scaling law relating the Sherwood to the P\'eclet numbers (\ref{sherwood}). Because we do not measure $\dot{m}$ directly but the grain size reduction rate $\dot{a}$ instead, we rather define an effective Sherwood-like number as:
\begin{equation}
\tilde{\mathrm{Sh}} = \beta_a \beta_b \, \frac{\rho_p a \dot{a}}{D c_0} .
\label{modifiedsherwood}
\end{equation}
For a spherical particle, for which $\dot{m}$ and $\dot{a}$ are simply related (Eq.~\ref{dotM2}), and setting the corrective factors $\beta_{a,b}$ to unity, both definitions of Sh and $\tilde{\mathrm{Sh}}$ coincide. Here, we not only wish to express this number with quantities we have direct access to, but also aim at accounting for the corrections we have discussed above. Similarly, the effective P\'eclet number writes
\begin{equation}
\tilde{\mathrm{Pe}} = \beta_a \, \frac{Ua}{D}
\label{modifiedpeclet}
\end{equation}
It can be directly estimated along each experimental run, also accounting for the radius correction. Plotting $\tilde{\mathrm{Sh}}$ as a function of $\tilde{\mathrm{Pe}}$ for all of our data clearly provides the expected increasing trend (Fig.~\ref{fig:shversuspe}). Data scattering is important, on the order of $30$\%, which is similar to what is displayed in Figure~\ref{fig:result}. For comparison to theory, $\tilde{\mathrm{Sh}}$ is computed with $U$ and $\dot{a}$ from their corrected expressions (\ref{Ucorrected}) and (\ref{dotacorrected}), setting the factors to the experimentally-determined averaged values $\beta_a = 0.921$ and $\beta_b = 0.988$, and where $a$ is deduced from $\tilde{\mathrm{Pe}}$ with (\ref{modifiedpeclet}). The agreement is quantitative, showing self-consistency with the fit of the theory in Figure~\ref{fig:result}.

\section{Conclusion}

We have investigated the dissolution of an almost spherical particle during its sedimentation, in the low Reynolds and high P\'eclet regime. We use butyramide particles sedimenting in aqueous solution so that the density contrast between the particle the solution is small, and thus the sedimentation velocity. The advantage of butyramide is that the density of the saturated solution is very close to the one of water, i.e. the dissolution does not affect the density of the solution.

The particle sediments in a squared tube, where a saturated butyramide layer is placed on top of a NaCl layer. The role of the top layer is to measure the sedimentation of the particle without dissolution and to have time to focus on the particle. The shape and the position of the particle are measured simultaneously by a camera attached to a translation stage. The particle is tracked in real time, and the translation stage moves accordingly to keep the particle in the field of the camera. 

We develop a simple model for a perfect sphere based on Stokes' law (hypothesis of low Reynolds number) and the mass transfer at low Reynolds and high P\'eclet derived in \cite{levichPhysicochemicalHydrodynamics1963}. We obtain a radius shrinking rate $\dot{a}$ which is constant in time, and only depends on the properties of the solid and the aqueous solution.  The position of the particle is a third order polynomial of the time $t$. In the experiment, we define an equivalent radius from the image of the particle. We find as expected by the simple model that  $\dot{a}$ is constant in time, and independent of the initial radius of the particle. Moreover, the theoretical and experimental results are consistent within $20\%$ without any adjustable parameter.

In order to obtain an even more quantitative agreement, we introduce two correction factors: one to take into account the non-sphericity of the particle in the evaluation of its volume and weight ($\beta_a$), and a correction of the density of the particle due to the inclusions of air bubbles inside the particle ($\beta_b$). The non-sphericity of the particle and the inclusion of air bubbles are visible on the images. These two correction factors are close to one ($\beta_a=0.921 \pm 0.002$ and $\beta_b=0.988 \pm 0.002$).
 With these corrections, both the radius shrinking rate deduced from the equivalent radius and the one deduced from the particle trajectory are in quantitative agreement with the corrected model. We discuss other correction factors, such as the correction of the drag due to the non-sphericity of the particle, the correction of the mass transfer due to the non-sphericity of the sphere and the finite solubility and non-isothermal effects in the dissolution of butyramide. We have shown that these corrections factors have a negligible effect in the present experiment, in contrast with $\beta_a$ and $\beta_b$. Finally, we have defined an effective Sherwood number $\tilde{\mathrm{Sh}}$ and an effective P\'eclet number $\tilde{\mathrm{Pe}}$, and we have displayed the corresponding curve, which show the $\tilde{\mathrm{Sh}} \propto \tilde{\mathrm{Pe}}^{1/3}$ scaling.


\section{acknowledgement}

We gratefully acknowledge X. Benoit-Gonin, A. Fourgeaud and L. Quartier for technical support and A. Limare for technical help and scientific discussions. N.H. and Y.C. have been funded by Chinese Scholarship Council Scholarship.

\bibliography{sample}

\end{document}